# Mitigating Botnet Attack Using Encapsulated Detection Mechanism (EDM)

**Maxwell Scale Uwadia Osagie[1*], C. I. Okoye[2] and Amenze Joy Osagie[1]**

[1]*Department of Physical Sciences, Faculty of Science, Benson Idahosa University, P.M.B 1100, GRA, Benin City, Edo State, Nigeria.*
[2]*Department of Computer Science, Enugu State University of Technology, P.M.B 1100, GRA, Benin City, Edo State, Nigeria.*

## ABSTRACT

Botnet as it is popularly called became fashionable in recent times owing to it embedded force on network servers. Botnet has an exponential growth of about 170, 000 within network server and client infrastructures per day. The networking environment on monthly basis battle over 5 million bots. Nigeria as a country loses above one hundred and twenty five (N125) billion naira to network fraud annually, end users such as Banks and other financial institutions battle daily the botnet threats. The most worrisome part of the botmaster's botnet is it propagation as an entity even when it is known to be large pool of malicious threats. The attacks leave end users (clients) to the risk of losing valuable credentials when connected to the affected infrastructure. It is on the above premise that this paper sort to expose the botnet method of propagation through proactive mechanism called Encapsulated Detection Mechanism (EDM) for botnet on Server Systems with further operations on conceptual framework, structural modules, usability and application of botnet. The mechanism uses one dimensional data stream evolutionary window approach of Distance Base Model (DBM) as an

_________________________________________________________________________________

*\*Corresponding author: E-mail: mosagie@biu.edu.ng;*

Outlier Analysis (OA). The Captcha, Username password and EDM Analyzer act as the front end of the data stream checker using Bot-Stream OutlieR Miner (B-STORM) algorithm and B-Exact Algorithm. The research work showed high level of data entering compliance efficiency on the server end network by neutralizing and mitigating botnet attack that falls short of the predefined data order within the networking signature.

*Keywords: Propagation; pool; botnet; networking; server and client.*

## 1. INTRODOCTION

The Currently improved information and communication technologies have enabled some unscrupulous individuals to cultivate systematic approach as to using the technology in committing all forms of illicit acts, create tensions and anxiety amongst organizations, firms, government agencies and individuals that rely solely on the use of network for day to day activity. Botnet is regarded as one of the most dangerous attack on a network due to the flexible nature at which it masquerades itself via server-end to people's computers and make such computers becoming bots (Zombies) [1]. The botnet uses an Internet Relay Chat (IRC) protocol designed for communication between two or more people online. Historical background has it that IRC protocol was designed in Finland by Jarkko Oikarinen in 1988 and in the late 80s through 90s it experienced tremendous patronage in networking and data communication [2]. Jarkko Oikarinen an Information Communication Technology (ICT) expert from the University Oulu, Finland, created IRC to replace the Multi-User Talk (MUT) program on the University OULUBOX [2]. The original idea for IRC creation over the years has vehemently been compromised thereby giving thousands of illicit hackers' opportunity to sporadically and perpetually unleash attack/terror on lawful client computer users due to the friendly nature of IRC protocol and server. The IRC is referred to as the main stream for multiple client's communication because it is the bridging block between two or more client computers who had agreed to chat over line link server [2,3]. The picture represented in Fig. 1 demonstrates connected systems using a single point server. The connected client systems share the resources provided by the server, while the robot represented by blue bot is said to be the affected server. Other connected robots with red bot are the affected client systems, and the target, as represented by the blue bot robot means all connected clients have been compromised.

Internet Relay Chat (IRC) was used to connect different chat rooms with the basic idea of exchanging messages which gained wider spread at the time it was created. Though, still in use but it has experienced a change in paradigm due to the new emergence protocol such as "I Seek You" as ICQ for short, Instant Messenger Protocol as AIM, and Messenger Protocol as MSN currently in use in networking [3]. These protocols have better orientation with Open System for Communication in Real Time (OSCAR). What the bot does in networking is to scan systems with little or no security control (vulnerability of systems). A full compromise of this system means the botmaster is in full control of the open channel. The waiting command and control (C&C) channel by a botmaster is the worrisome threat in the server platform and can execute over 200 commands at a time. Botnet as a trend in computing world remains a terror and either ways it presents itself should be discouraged by both server and client operators and along with end users [2,3]. The bot-master uses botnet via server end in making client computers/systems becoming bots (Zombies) and thereafter makes such systems subject to the full control of a bot-master. The activity of a bot-master if in control can create service denial within connected systems using the affected server and this creates room for extracting useful information from the affected (zombies) computer system [4,1]. According to [5] botnet is a large pool of compromised host that are controlled by a bot-master. They explained that recent botnets uses the Internet Relay Chat (IRC) server as their Command and Control (C&C) server for controlling the botnet. Bot-master can disperse command to its botnet by the use of the IRC C&C channel. It has been showed that most botnets use the IRC for C&C processes. However, the traffic among bots, the C&C server and the bot-master can be considered as legitimate traffic because it is hard to distinguish it from normal traffic. According to [5], on a report by CipherTrust states that the constant propagation of botnet in scope and size amounts to about 172,000 bots and this statistics is on a daily recruitments; this means that about 5

**Table 1. Survey on Nigeria internet users**

| Year | Internet users** | Penetration (% of pop) | Total population | Non-Users (internetless) | 1Year user change | 1Year user change | Population change |
|---|---|---|---|---|---|---|---|
| 2016* | 86,219,965 | 46.1 % | 186,987,563 | 100,767,598 | 5 % | 4,124,967 | 2.63 % |
| 2015* | 82,094,998 | 45.1 % | 182,201,962 | 100,106,964 | 8.4 % | 6,348,247 | 2.66 % |

*Source:[6]*

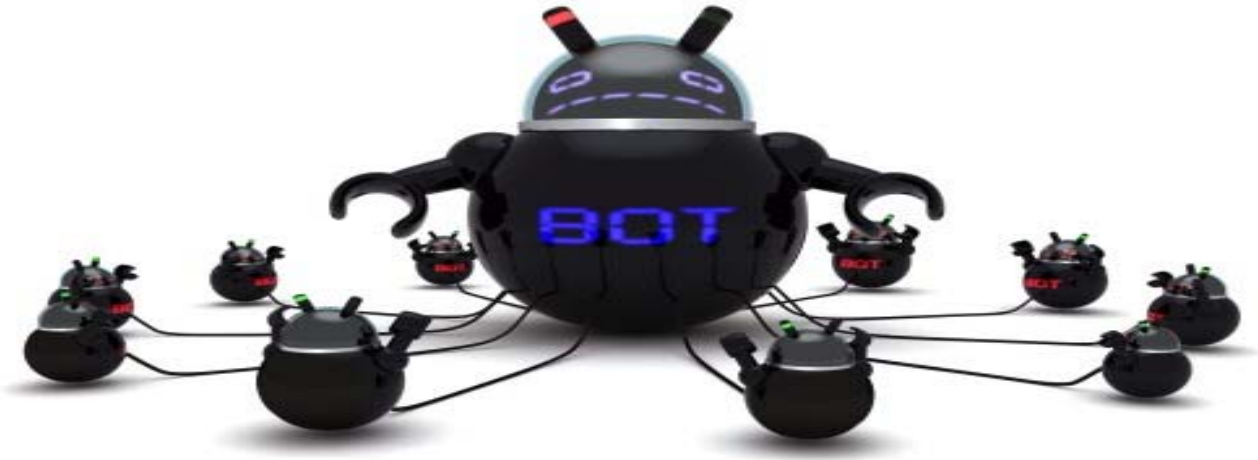

**Fig. 1. Pictorial view of a Botmaster's bot**

million new bots are released every month. This is a worrisome proportion to any advanced nations much more developing nations such as Nigeria. The Same article by [5] also made known Symantec report that the number of bots observed in a day is 30,000 on the average. The total number of bot infected systems has been measured to be between 800,000 - 900,000, a single botnet comprising of more than 140,000 hosts was found in the wild and botnet driven attacks have been responsible for single Distributed Denial of Service (DDoS) attacks of more than 10Gbps capacity [5].

A cross examination of Table 1. above unveil the progress made so far by Nigeria and Nigerians in embracing the reality of what internet has to offer. The years captured is from 2015 to 2016. As at 2016 the internet penetration within Nigeria spectrum or population as at July was seen to be 86,219,965 representing 46.1 percent of the total population of 186,987,563. This is a worrisome figure or percentage for a country that wishes to stay safe from the ravaging nature of botmasters. What this mean is that the awareness of internet usage is still less than half of the total population. The conceptual agreement here is that the figure above involves desktop, laptops, palmtops, and all other handheld devices. A further classification might reveal lesser percentage of those actually involved in real time and business computing not the pleasurable (Social Media) users. However, in responding to the questions above some logical deductions are made from certain principles and they could be classify as moral but not lawful.

## 2. RELATED LITERATURES

The criteria for judging the deficiency of botnet detection present systems varies along different line adopted by authors/researchers into botnet threats. The propounding notes in these methods are the identification and protection of server network from bot as claimed by few mechanisms. The most worrisome is simply the fact that the trend has continued to make soft landing in data invasion even when these mechanisms thus far had since been seen as the bedrock to none-invasion of secure data [5]. There are several reports from print and broadcasting media stating clearly how stealing of unauthorized data from server or client systems have become and this has led to millions of dollar unaccounted for on the part of the financial institutions (Banks) around the globe. According to [7], the Nigeria Inspector General of Police (IGP) Ibrahim Idris attests to the fact that systems networking (cybercrime) has exponentially increased in recent times, he made this known via an Interpol Cybercrime Training for Practitioner Investigators from Africa Countries (ICTPIAC) held in Abuja. The report, which was widely published states that the Nigeria Police Force has swung into action by establishing a cybercrime unit to curb the increase. In another report by [8] via a vanguard newspaper in July 27, 2016, reported that the Microsoft Digital Crime Unit (DCU) had revealed that Malwares cost the global economy $3 trillion annually. It is also postulated that in every seconds, up to 12 people becomes victim of cybercrime, a situation considered both of national and international disgrace. This chapter takes swipe into these areas and helped provide an insight into the system proposed and how best it can solve (in part or full), the problem of immerse increase in botnet threat.

The most common navigating spectrum of botmaster is the IRC, HTTP and Peer2Peer network hijacking. However, long before the Peer2Peer was introduced in the year 2007, the IRC bot had long seen some detection mechanisms used by server operators in curbing the botmasters excesses [9]. Other detection mechanisms such as NetFLow, DNS group activity traffic, Intrusion Detection System (IDS), and PROVEX as captured by [10,5,11,12] focuses on the IRC design of a single point host. [13] did a survey on HTTP bot and presented a published survey article on botnet detections and unveil via the widely published article that HTTP bot uses TCP as the driving force of its propagation but this focus was on HTTP bot. Similarly, [14] also did an empirical study on botnet on large scale network called "Bot-meter". According to [15], botnet can be detected by supervised learning methods, such as classification and regression, or by unsupervised learning methods such as clustering. In a method articulated into three (03) dimensions. A three most popular clustering Algorithm in hybrid model combined with decision tree to classify malicious flows. The above method by [15] saw an evaluation of $DR = TP/TP+FN$ where TP is the number of malicious flow correctly identified as normal flow, while FN means numbers of malicious flow that are incorrectly identified as normal flow and the DR represents the detection rate. At the end of the exhaustive examinations, it was revealed that K-Means could help identify malicious flow. [16] In a research titled Management and Security in the Age of Hyperconnectivity, the authors revealed that the emergence of Self-Organizing Network (SON) has helped increase the proliferation of Network threats such as botnet and also helped solve this loophole, a concept known as Network Element Virtual Temperature (NEVT) to create the needed stability in the Network Elements (NE).Similarly, [17] proposed a DeDroid detection technique for mobile botnet, which focused on static code analysis, considering permissions and API calls. The static analysis provides a lightweight approach as compared to the dynamic analysis.

For clarity purposes, the different detection techniques used thus far in combating botnet were adequately examined in the reviewed but of most significant in all the reviewed techniques is the fact that the operational goal is to detect the bot within a given system or architecture on a single or double navigation methods use in gaining access. Another understanding gained from the research thus far is the fact that all botmasters have one common goal and which is to invade the system architecture, then removes valuable credentials. Though, the method of propagation or penetration differs, they still have similar understanding in the manner to which the system theft is accomplished and these vary from stealing of credentials to causing denial of service (DoS) to the entire network. The constant approach by botmasters into network servers gave the blueprint to the solution proposed by this study which is believed to be the most proactive detection mechanism ever proposed. This is seen from the unique features of the "fight back", and captcha mechanism along with the embedded systematic approach to combining all other mechanisms such as IRC, HTTP and Peer2Peer bot into one entity (EDM).

## 3. BOTNET TOPOLOGY

Categorizing botnet in networking is not as easy as many would envisage. Topology is far more than naming inbound/outbound threats and it is considered to be more tedious than cracking the gene of corruption within Nigeria politics. This is because botnet as defined by [5] is a large pool of compromised host that are controlled by a botmaster using the command and control (C&C) channel. This means that the compromised nature of the host attack cut across not one single malware and if a trace is to be carried out, the compromised malwares coming together as a pool must be critically examined [18].

Fig. 2 unraveled the different types of topology as explained by [19], from the chart, the IRC is having 38.2% acceptability in the community of botmasters thereby leaving the HTTP as the closest rival in full penetration of 29.1% in botmaster's business. The Peer2Peer, which is also considered as new platform for botmaster has 2.3% impact while others are categorized to have wide effect of 30.5%. These statistics as at 2011 made open the amount of attack gear towards server using Internet Relay Chat (IRC) and other attacks capable of bringing down the functionality of other topologies. Since, the introduction of IRC as captured by [2] and it collaboration with botmaster several botnet tools it would be difficult to pin it down to one category. Nevertheless, [1] captured four of the below listed categories of botnet and the six (6) captured by [20]. Detailed survey unveiled other kinds of bot tools, and these also include the "random bot". Below are the various bot types design structures.

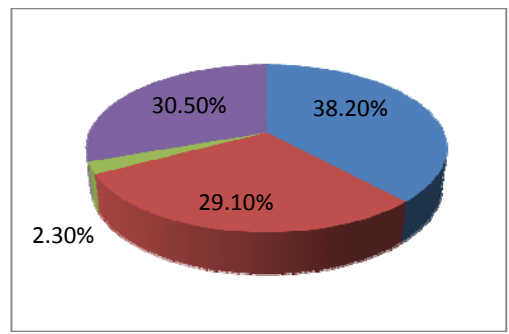

**Fig. 2. Botnet command and control (C&C) topology**
*Source: [19]*

### 3.1 Localized Bot

The localized bot is considered as new trend [21] but this is being considered as the next generation of bot because botmasters are currently writing scripts and malwares in programmable language that could navigate some operating systems platform. Well, bot is not new to Microsoft window's operating systems.

### 3.2 Peer2peer Bot

At the time botnet found it root on server's operational scheme, history never revealed how it will metamorphose into current trend categories [21]. The IRC bot is a server oriented protocol that has its activity centre on Server Single Point Hosting (SSPH) which made it possible to detect partial activities of botmaster due to the centralization of the server network. The centralization of the server network made it mandatory for client system to come into the server with a Single Point Entry (SPE) [22]. This enable the Server Entry Port (SEP) to identify different Domain Name System (DNS) and IP clustering classification based on pre-registered information available to the server DATASET [15]. To this end, botmasters quest to seek alternative to server intrusion began to yield fruit from the discovering of peer2peer bot topology [21]. The effrontery to be in charge was at this point considered nothing because shutting down the server would lead to a collapse in the operational paradigm, loss of money, time, and energy in the bostmaster's end [20].

- ➢ **Centralized Peer2Peer Computing Structure:** According to [23], the botnet structure is subdivided into centralized, decentralized and Hybrid Peer2Peer. The centralized Peer2Peer Computing Structure captures the server network paradigm in much better way and it involves client systems with the use of single port entry (SPE) as captured in Fig. 3.
- ➢ **Decentralized Peer2Peer Computing Structure:** The decentralized architecture/design is a comprehensive topology of a network that takes care of all connected systems through different host/port. Thus, all Peer2Peer systems with DNS identity do have access to the network via different means unlike the centralized topology that identify client systems through single host (client-server network) and this is demonstrated in Fig. 4.
- ➢ **Hybrid Peer2Peer Computing Structure:** A comparison of the behavioral pattern of the centralized Peer2Peer Computing Structure and the Decentralized Peer2Peer Computing Structure elucidates the dichotomy in both practical and theoretical perspective. However, the restructuring of centralized and the non centralized structure in forming one indivisible and formidable structure is the Hybrid Peer2Peer Computing Structure [24]. However, it is difficult to say that there are computing industries or information technology companies using the structure captured in Fig. 5. Also of great concern, is the enormous pool of compromised botnet tools from different botmasters [25].

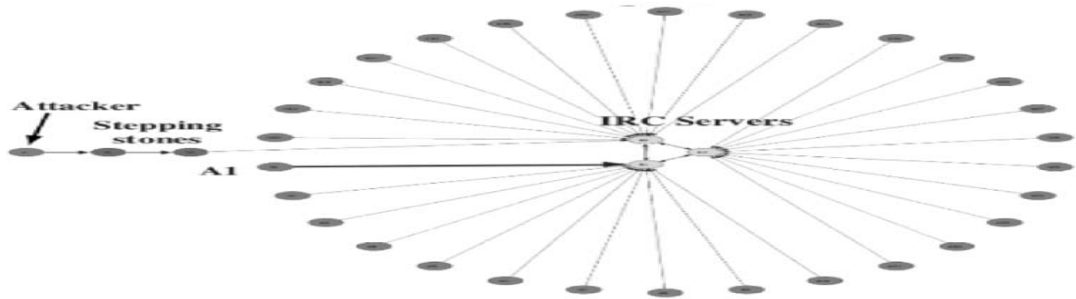

**Fig. 3. Centralized Per2Per Computing Structure**
*Source: [24,23]*

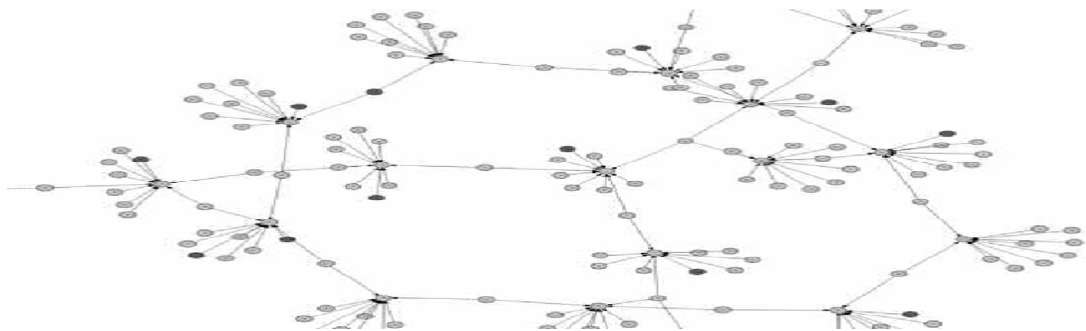

**Fig. 4. Decentralized Peer2Peer Computing Structure**
*Source: [24]*

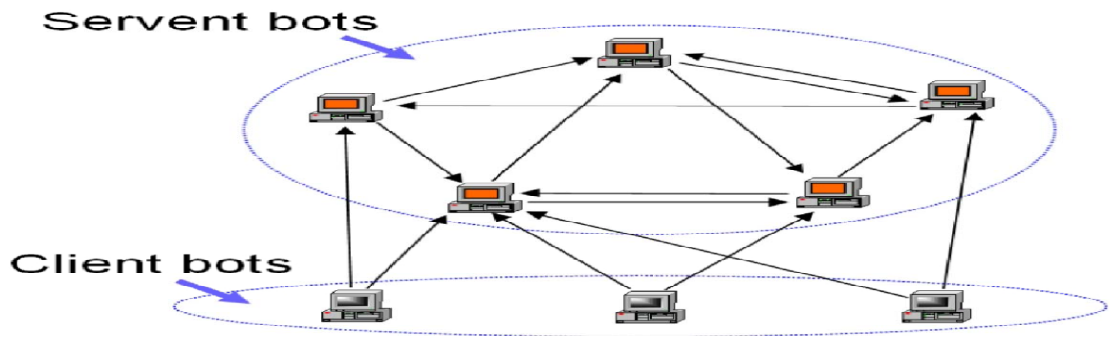

**Fig. 5. Hybrid Per2Per Computing Structure**
*Source: [24].*

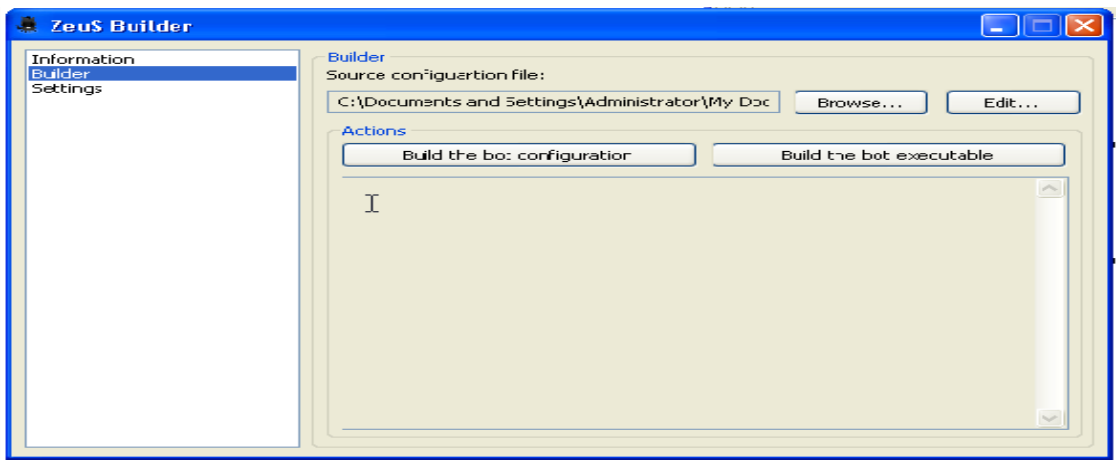

**Fig. 6. Zeus (Zbot) builder computing platform**
*Source: [1]*

### 3.3 HTTP Bot

The HTTP bot has the embedded ability to cause serious DoS attacks and being a pool of malware, it circulates worm by propagating it randomly, scouting and scanning without destructing the client system browser which in this case is the host. The negative and most disheartening phenomenon about the HTTP bot is that it does not connect to a host [26,27]. Rather than the HTTP connecting to a host like other botnet methods, it follows web pages depending on the method of propagation. Hence, making it difficult to trace or track by anti-botmasters. The HTTP bot has succeeded in harmonizing and aligning with the normal traffic [28], making its identification an enormous task and the technicalities involved require a team work [29].

Zeus/Zbot is a compromised malware classified as Trojan, a malware package with multiple functionalities such as running on Microsoft window and same time carrying out other act of criminalities in more deceitful manner in network environment and thereafter steals personal credentials from the compromised system [30]. The stolen items cut across bank details and information of the victim's system. The Zbot as seen by Fig. 6 does the stealing of victim's information in much pretty format by perfectly using the "Keystroke Logging and Form Grabbing". The proliferation of Zbot is through Drive-By-Downloads (DBD). Another area where the proliferation takes place is through phishing [31,1].

### 3.4 Random Bot

The random botnet/bot posses both undermined and none undermined malware activities with its comparative advantage over connected systems, the comparative advantage is on the fact that the spread of bot is patternless within connected systems. The destructive syndicate in botnet structure or topology in a well organized system platform has become a source of worry amongst server operators [32,33].

### 3.5 Spy Eye Attack

Spy eye (Spyeye) as captured in Fig. 7 posses some attributes considered not compactable with other malwares. This is because, similar tools such as Zbot/Zeus share same functions in its operational characteristic. The supremacy of SPYEYE over ZBOT is on its comprehensive design that enables its delete Zbot when found together within a system of internetworking platform [34,1,35,36]. The controversy on who is in charge has been the most dangerous trend amongst Botmasters. Spy eye being a well established design with good architecture or topology in navigating systems on networking platform or the internet, reserved the right and obligation to superlatively control Zbot.

### 3.6 Torpig

Every botmaster knows the importance of traffic utilization and will stop at nothing till the aim is achieved. The diversification of different computing gadgets has simplified the work of an average botmaster, different gadgets, though maintained set of coherent and comprehensive security standard, but lack of measurement in its software and hardware apparatus has increased the proliferation of attack from botmasters in no little measures [37]. Torpig as shown in Fig. 8 is a tool capable of compromising the functionality of network through mebroot toot kits, a highly endow Trojan horses capable of stealing credentials.

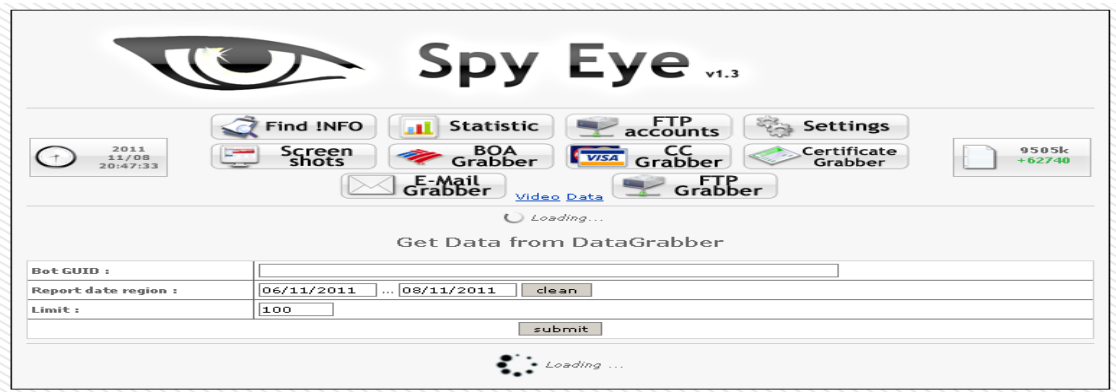

**Fig. 7. Spy eye Attack builder interface [20]**

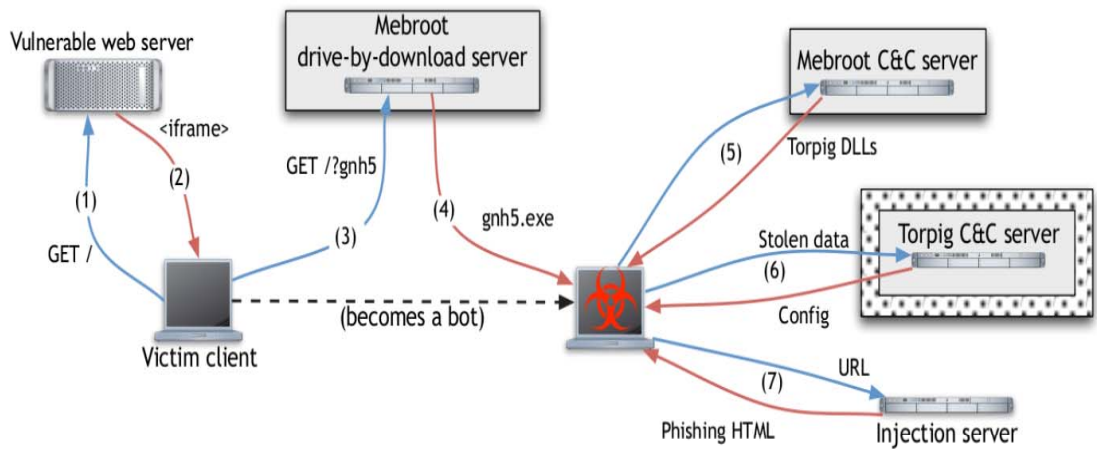

**Fig. 8. Torpig network infrastructure**
*Source: [20]*

## 4. MODEL APPROACH FOR BOTNET MITIGATION

To actualize the proposed model (EDM) and to capture data (bot) that falls short of the pre-defined order within the network window, the design was model using one dimensional data stream evolution model proposed by [38]. This model was considered suitable because the approach best described data tagged to be Outlier (bot/Botnet) and considered being harmful to the system if allowed to stay. Inlier data are data considered to be legitimate with full authorization to navigate the server/network space.

A critical examination into the operational function of Fig. 9 shows stream movement within segmented classification. The operation paradigm of Distance-Based Outlier could be examined from Fig. 9 which shows data evolution windows. The first window at time 1 –time 18 shows that o9 and o11 are inliers because they could be seen as having four (4) neighbors in its classification (o5, o10, o14, o15). o11 has four (4) neighbors (o3, o4, o6, o13). Considering that o9 has three succeeding neighbors it is known to be safe inlier, while o11 is not known to be a safe inlier. At time 22, o9 is still an inlier while o5 has gone into extinction (waste), but o9 still has three succeeding neighbors. This automatically makes o11 an outlier. Object o3, o4, and o6 have all gone into extinction (expired) because only one neighbor is attributed to them [38]. Based on Fig. 9 the proposed model was formulated as seen below.

## 5. PROPOSED MODEL (EDM)

The design in Fig. 10 is a centralized networking infrastructures linked together by a common goal. Though, each infrastructure has distinct role but the collective and operational synergy brings about operational output. The client systems are handheld devices capable of functioning under the guideline Principles of Open System Interconnection Model (OSI Model) and Secure Socket Layer (SSL) for information exchange and authentication respectively. The Server Side is made of the information requires by the client system and being the platform for access it is usually the critical part of a network that an attacker which to log into with the sole purpose of making it bane. The server mechanism is the model designed to sort out entering protocols against the required type through SSL approach.

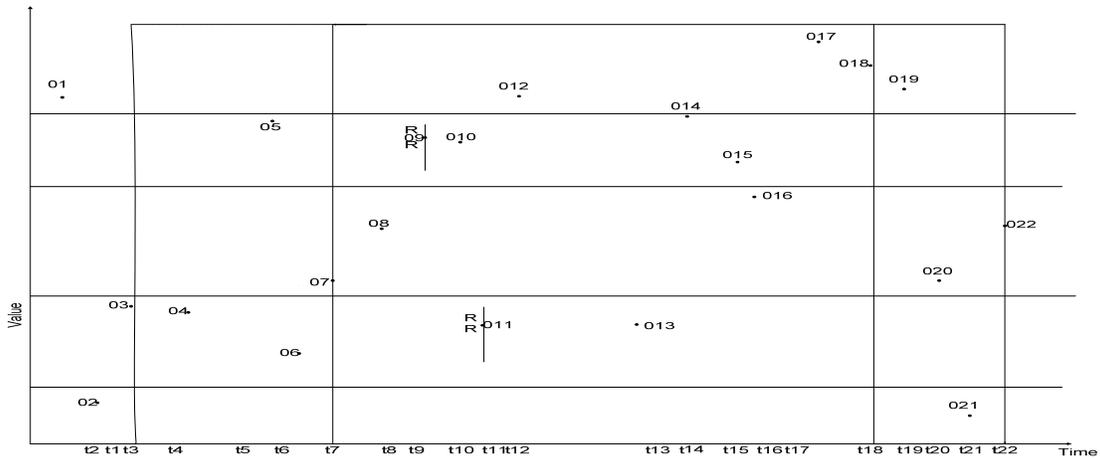

**Fig. 9. Evolution of one dimensional data stream**
*Source: [38]*

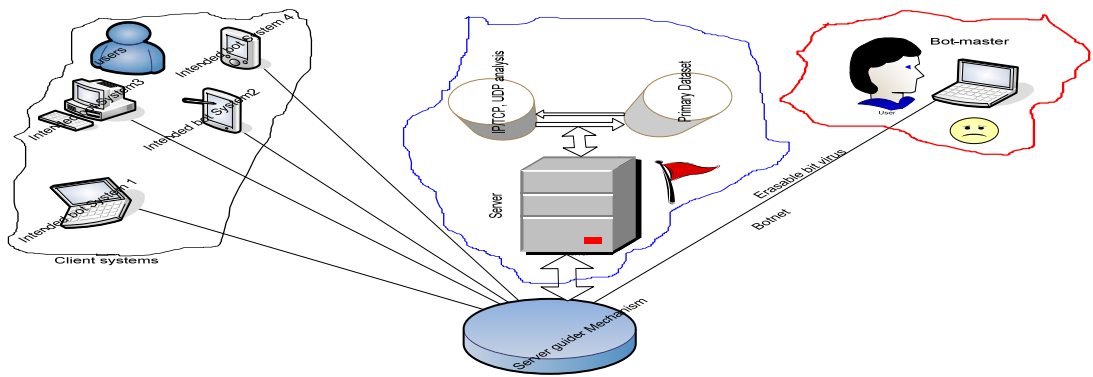

**Fig. 10. An encapsulated detection mechanism for botnet on server end**

## 5.1 Design Structure in Details

Critical examination into some mechanisms discussed in the related literatures shows that they are classified to be reactive in operations and taking care of botnet lies at the method of same propagation approach. The point enveloped in Fig. 11 is an encapsulated model in the server side with three segmented guide SSL authentication created in the entry layer and as well analyze, detect and fight back any bot or botnet that failed the EDM integrity at the functional layer of the EDM server. The system is designed in three modules with sub modules working as an entity in the actualization of server protection. The three modules and sub modules are outlined below.

### A. User layer

This layer has the enduring process of all legitimate users who at a point made and synchronized confidentiality with the system design for onward recognition (handshake or signaling)

### B. Authentication layers

1. Captcha: This eradicates suspicious entry on the system and then creates integrity and assigned privileges to legitimate users on the network server

2. Username and Password: The system grants access to predefined and registered users via this segment

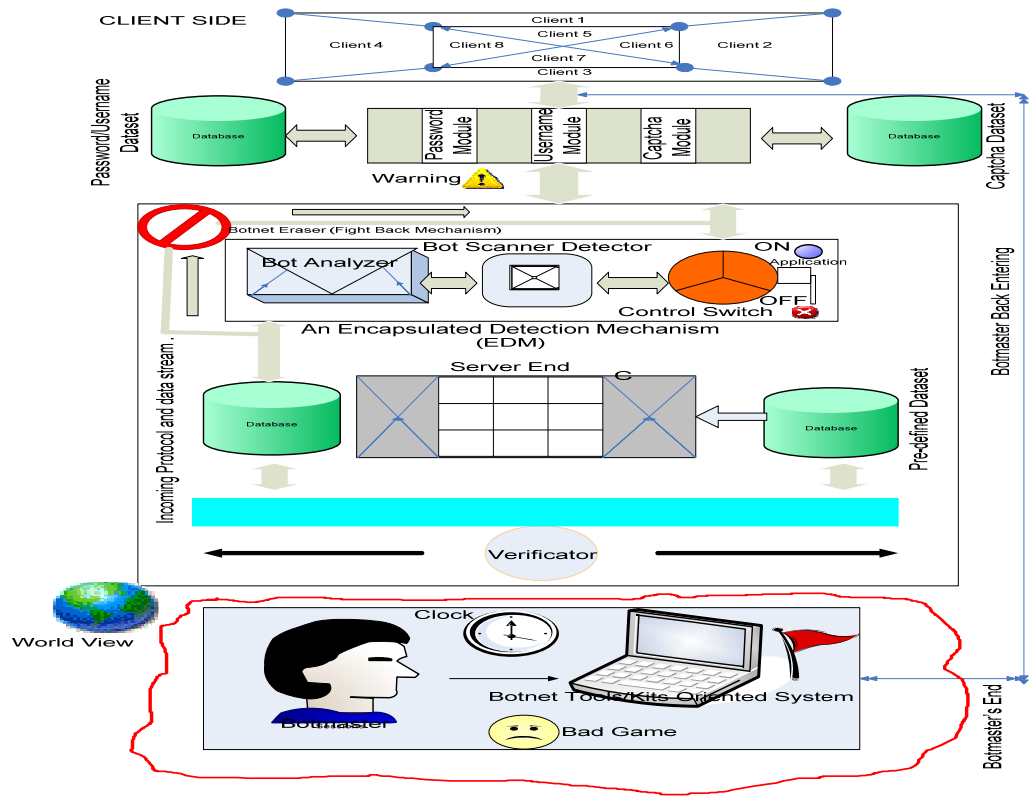

**Fig. 11. Architecture of an encapsulated detection mechanism for botnet on server end**

**C. Analyzer**

1. Encapsulated detection mechanism analyzes the data window movement in one dimensional array and there after verifies the movement against the predefined dataset in the server window. The mechanism has bot scanner, analyzer and a verification module, these sub mechanisms work as an entity in stimulating fight back against data movement not conforming with the lay down data predefined order

2. Bot scanner is a logical mechanism that goes through input data window, stream and check against unwanted data that may have found itself into the system server by way of futuristic propagation

3. Bot analyzer and verification module is a double process scheme for terminating a protocol action through the synchronization of data found to be outside the scope of the predefined dataset.

Fig. 12 is a demonstration of the design EDM with e-Banking application. This authenticated interface platform design to distinguish legitimate users from programmable machines such as robot and other automation. The Captcha requires user to type in the codes generated by the system and thereafter redirect to another authentication page of username and password which further carryout some data entry fundamentals before granting access to the EDM.

The EDM is a mechanism design to verify accessed device protocol against bot propagation in the form of HTTP bot, P2P bot and IRC bot hence Fig. 13. The EDM checks for bot on the client/server systems using the B-Storm and B-Exert algorithm through distance base outlier. It features involves the automatic blockage of such when found and send and instantaneous malware via same link to which the bot came through to the network or server.

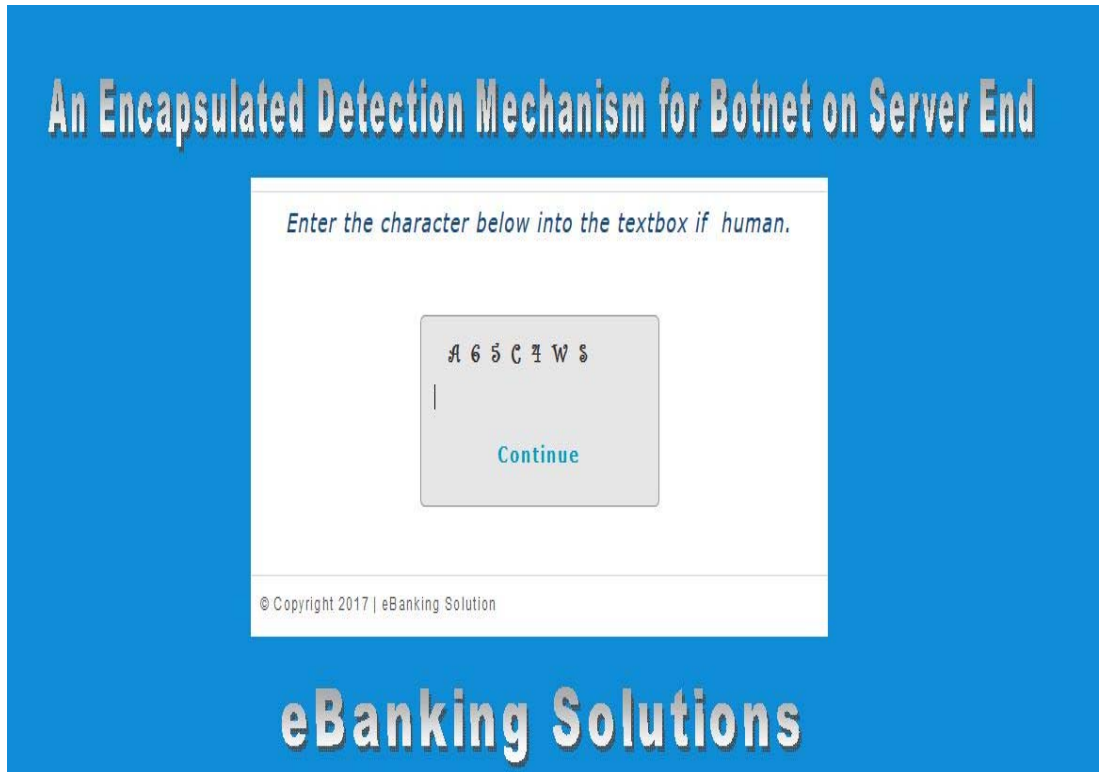

**Fig. 12. Encapsulated detection mechanism application interface**

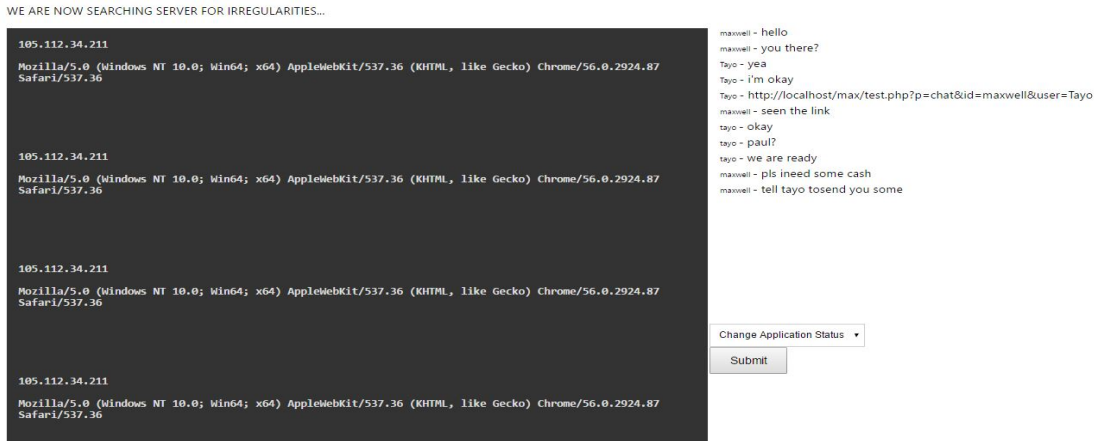

**Fig.13. EDM on active state**

## 6. SURVEY RESULT

The chart seen in Fig. 14 is a primary data that captured one of the factors increasing botnet propagation in recent times. From the frequency and percentage distributions in Table 3 where 50% of the respondents agreed that staff within the financial institutions contributes to the increasing rate of botnet and the justification to this data is the report by the IGP and the increasing rate of fraud stars ("Yahoo Boys") collaborations with banking staff.

**Table 2. Staff factor in botnet increased**

|  |  | Frequency | Percent | Valid Percent | Cumulative Percent |
|---|---|---|---|---|---|
| Valid | 1 | 15 | 13.9 | 13.9 | 13.9 |
|  | 2 | 6 | 5.6 | 5.6 | 19.4 |
|  | 3 | 33 | 30.6 | 30.6 | 50.0 |
|  | 5 | 54 | 50.0 | 50.0 | 100.0 |
|  | Total | 108 | 100.0 | 100.0 |  |

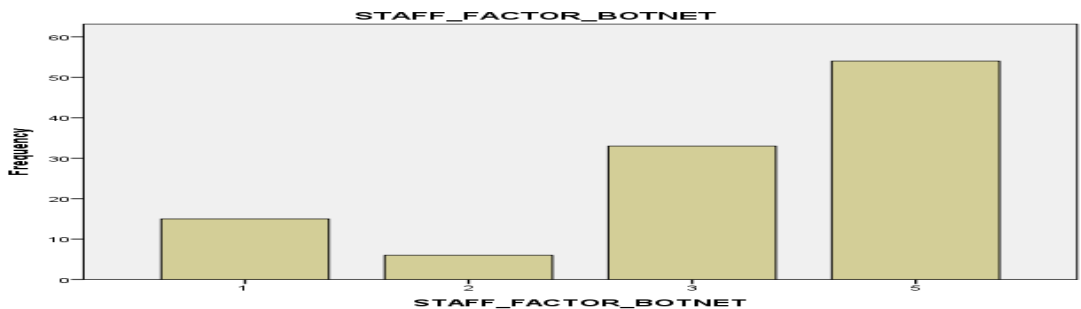

**Fig. 14. Chart representing staff role in botnet propagation**

**Table 3. Call for new security measure in curbing botmaster's botnet**

|  |  | Frequency | Percent | Valid percent | Cumulative Percent |
|---|---|---|---|---|---|
| Valid | 1.0 | 7 | 6.5 | 6.5 | 6.5 |
|  | 2.0 | 6 | 5.6 | 5.6 | 12.0 |
|  | 3.0 | 32 | 29.6 | 29.6 | 41.7 |
|  | 4.0 | 59 | 54.6 | 54.6 | 96.3 |
|  | 5.0 | 4 | 3.7 | 3.7 | 100.0 |
|  | Total | 108 | 100.0 | 100.0 |  |

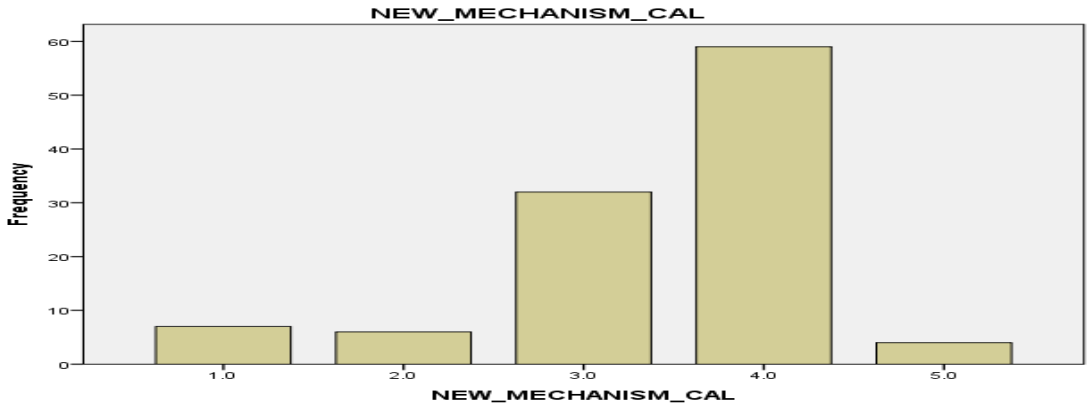

**Fig. 15. Chart for call for more defensive and proactive mechanism**

Table 2 shows other primary and secondary data on botnet, further increasing the propagation of the survey which therefore, sort to known if there are needs for a secure layer defensive approach to the operational status of Nigeria financial institutions networks and applications hence the Table 3 and Fig. 15 which clearly supports the need for not just defensive mechanism against botnet but a proactive defensive mechanism that could take back the fight to the propagation machine and thereafter force it out of server end network. However, 6.5% and 5.6 % in total of 12.1% of the respondents failed to support the call but 54.6% gave reasons why proactive mechanism must be introduce to curb the growing trend of botmaster's botnet and return income on investment on the part of the service providers and as well create confidence on clients.

## 7. RESULT AND CONCLUSION

The eradication of botnet using Distance Base Model Outlier Analysis, B-STORM and B-exert algorithm has demonstrated that botnet propagation prevention or detection could not only be said to be reactive within server as seen by other mechanisms examined. This paper unveils a new dimension in botnet mitigation using the same approach to which the botnet gained access to the server. From the Captcha introduction which verifies legitimate users as well as username password authentication and the EDM synchronization. The EDM scan entering protocols, carry out further checks on the validity of the entering data by analyzing and double verifying against pre-defined dataset through the above algorithms before responding in brutal manner. These are evidences to the immerse contribution to botnet prevention and eradication.

Since the introduction of internet relay chat (IRC) in1990s botnet has increased in both scope and sized. This can be seen from the statistical view of botnet propagation as captured by [5] as well as botnet penetrations in banking sector showed in Figs. 14 and 15. Botnet as captured in its definition is a group of malicious tool working as an entity within network domain. It is quite difficult to say the tool is for stealing only, it usage cut across distributed denial of service (DDoS), network terrorism, collection of vital or classified credentials, phishing, hijacking of entire process of network server etc. The proposed system has not only demonstrated strong will in stopping the Botmaster trend but has the structural strength of accounting for every data movement within the network domain. Preceding the survey output many network users sees network problem within financial institutions as normal or a particular machine working in reverse but this paper has exposed in details that this is not always true because most network failures could be attributed to the ravaging botmaster's botnet with the sole purpose of causing hardship on the part of clients and the financial institutions in general. The EDM embedded features is classified to be proactive and brutal in fighting back botnet via encapsulated malwares within network.

## ACKNOWLEDGEMENT

This research paper wouldn't have been possible without the scholarly contributions from distinguished Professor C.I. Okoye, Dr. M. Omoruyi, and Mr. M. A. John-Otumu and the financial support from Benson Idahosa University, Benin.

**COMPETING INTERESTS**

Authors have declared that no competing interests exist.

**REFERENCES**

1. Bu Z, Bueno P, Kashyap R. The New Era of Botnets; 2010.
Available:http://www.mcafee.com/in/resources/white-papers/wp-new-era-of-botnets.pdf
2. Susan CH. Computer-mediated communication on the internet, Indiana University. Annual Review of Information Science and Technology; 2002.
3. Robert E. Creation of Internet Relay Chat Nicknames and Their Usage in English Chat room Discourse. *Linguistik online* 2011;50:6/11.
Available:http://www.robertecker.com/hp/research/leet-converter.php
(Accessed November 8, 2011)
4. Ahmad K, Rosli BS, Muhammad S, Syed AAS. Botnet detection techniques: Review, future trends, and issues. Irfan AWAN2, Nor Badrul ANUAR1. (1Faculty of Computer Science and Information Technology, University of Malaya, Kuala Lumpur, Malaysia) (2Department of Computer Science, University of Bradford, Bradford BD7 1DP, UK). E-mail: ahmadkarim@um.edu.my Received Aug. 31, 2013; Revision accepted Jan. 23, 2014; Crosschecked Oct. 15, 2014 Journal of Zhejiang University-SCIENCE C (Computers & Electronics) ISSN 1869-1951 (Print); ISSN 1869-196X (Online) www.zju.edu.cn/jzus; www.springerlink.com E-mail: jzus@zju.edu.cnKarim et al. / J Zhejiang Univ-Sci C (Comput& Electron) 2014; 15(11):943-983
5. Choi H, Lee H, Lee H, Kim H. Botnet detection by monitoring group activities in DNS traffic. In Computer and Information Technology, 2007.CIT 2007. 7th IEEE International Conference on. IEEE. 2007; 715-720.
6. Internet Live Stats. Nigeria Internet Users *(www.InternetLiveStats.com) Elaboration of data by* International Telecommunication Union (ITU), World Bank, and United Nations Population Division; 2016. Retrieved September 16, 2016.
7. Yahaya F. Soaring Cybercrime Alarming says IG; 2016.
Available:http://thenationonlineng.net/soaring-cybercrime-alarming-says-ig/
8. Aginam E, Elebeke E. Nigeria Loses N127bn annually to cybercrime-Adebayo Shittu, Vanguard News; 2016.
Available:www.vanguardngr.com/2016/07/ngria-loses-n127bn-annually-cybercrime-adebayo-shittu/
9. Binkley JR, Singh S. An algorithm for anomalybased botnet detection. Proc. USENIX Steps to Reducing Unwanted Traffic on the Internet Workshop. 2006;43-48.
10. Cisco N; 2008.
Available:www.cisco.com/en/US/products/ps6601/products_ios_protocol_group
11. Grizzard JB, Sharma V, Nunnery C, Kang BB, Dagon D. Peer-to-peer botnets: Overview and case study in Proc. 1st Workshop on Hot Topics in understanding Botnets; 2007.
12. Christian RY, Christian JD. PROVEX: Detecting botnets with encrypted command and control channels. University of Applied Sciences Gelsenkirchen, Institute for Internet Security, Germany; 2014.
13. Chaware SP, and Bhingarkar S. A Survey of HTTP Botnet Detection. International Research Journal of Engineering and Technology (IRJET) e-ISSN: 2395-0056 2016;03(01).
Available:www.irjet.net p-ISSN: 2395-0072
14. Wang T, Hu X, Jang J, Ji S, Stoecklin M, Taylor T. BotMeter: Charting DGA-Botnet landscapes in large networks. In Distributed Computing Systems (ICDCS), 2016 IEEE 36th International Conference IEEE. 2016;334-343.
15. Mai L, Park M. A comparison of clustering algorithms for botnet detection based on network flow. In Ubiquitous and Future Networks (ICUFN), 2016 Eighth International Conference on IEEE. 2016; 667-669.
16. Badonnel R, Koch R, Pras A, Drašar M, Stiller B. (Eds.). Management and Security in the Age of Hyper connectivity: 10th IFIP WG 6.6 International Conference on Autonomous Infrastructure, Management, and Security, AIMS 2016, Munich, Germany, June 20-23, 2016, Proceedings. Springer. 2016;9701.

17. Karim A, Salleh R, Khan MK, Siddiqa A, Choo KKR. On the analysis and detection of mobile botnet applications. Journal of Universal Computer Science. 2016;*22*(4): 567-588.
18. Rajab M, Zarfoss J, Monrose F, Terzis A. My Botnet is bigger than yours (maybe, better THAN yours) : Why size estimates remain challenging. In USENIX Workshop on Hot Topics in Understanding Botnet; 2007.
19. Emre Y. A literature survey about recent botnet trends. GEANT 3 JRA2 T4: Internal deliverable ULAKBIM; 2011. Available:Turkey emre@ulakbim.gov.tr
20. Brett S, Marco C, Lorenzo C, Bob G, Martin S, Richard K, Christopher K, Giovanni V. Your Botnet is My Botnet: Analysis of a Botnet Takeover Department of Computer Science, University of California, Santa Barbara; 2009. {bstone,marco,sullivan,rgilbert,msz,kemm, chris,vigna}@cs.ucsb.edu
21. Narang P, Hota C, Sencar HT. Noise-resistant mechanisms for the detection of stealthy peer-to-peer botnets. Computer Communications; 2016.
22. Jing J, Helal AS, Elmagarmid A. Client-server computing in mobile environments. ACM computing surveys (CSUR). 1999;31(2):117-157.
23. Matthew S, Igor I. Detection of peer-to-peer botnets, UvA supervisor: Dr. Ir. C. DE LAAT, SURFnet supervisor: Ir. R. SPOOR, SURFnet supervisor: Dr. W. BIEMOLT. Research Project 1 Master of Science Program Academic year 2007{2008 Acknowledgement. Amsterdam, February 4, 2008.
24. Dittrich D, Dietrich S. Command and control structures in malware.;login: publication; 2007. Available:http://www.usenix.org
25. Warmer M. Detection of web based command & control channels; 2011.
26. Venkatesh KG, Anitha R. HTTP botnet detection using adaptive learning rate multilayer feed-forward neural network. Department of Mathematics and Computer Applications PSG College of Technology, Coimbatore, India; 2012. Available:g.kiruba@gmail.com, anitha_nadarajan@mail.psgtech.ac.in
27. Masud MM, Al-khateeb T, Khan L, Thuraisingham B, Hamlen KW. Flow-based identification of Botnet traffic by mining multiple log files. In: Proceedings of the International Conference on Distributed Framework & Application, Penang, Malaysia; 2008.
28. Guofei Gu. BotMiner: Clustering analysis of network traffic for protocol and structure independent botnet detection, Proceedings of 17th Conference on Security Symposium, ACM Digital Library. 2008; 139-154.
29. Wang B, Li Z, Li D, Liu F, Chen H. Modeling connections behavior for Web-based bots detection, In: 2nd IEEE International Conference on e-Business and Information System Security (EBISS) - 2010, Wuhan. 2010;1-4.
30. Binsalleeh H, Ormerod T, Bouhtouta A, Sinha P, Youssef A, Debbabi M, Wang L. On the Analysis of the Zeus Botnet Crimeware Toolkit. In: Proceedings of the IEEE Eighth Annual Conference on Privacy, Security and Trust,PST, Aug 17-19, 2010, Ottawa, Canada; 2010.
31. Jim FJ. Hackers steal U.S. government, corporate data from PCs. Reuters; 2007. Retrieved 5 September, 2016
32. Cooke E, Jahanian F, McPherson D. The zombie roundup: Understanding, detecting, and disrupting botnets. Steps to Reducing Unwanted Traffic on the Internet Workshop (SRUTI '05), Cambridge, Massachusetts, USA; 2005.
33. Abbes T, Bouhoula A, Rusinowitch M. Protocol analysis in intrusion detection using decision tree. In: Proceedings of the IEEE International Conference on Information Technology: Coding and Computing (ITCC, 04), 2004;404-408.
34. Karasaridis A, Rexroad B, Hoeflin D. Wide-scale botnet detection and characterization. In: First Workshop on Hot Topics in Understanding Botnets (HotBots'07), Cambridge, MA; 2007.
35. Xu T, Dake H, Luo Y. DDoS attack detection based on RLT features, proceedings of international conference on Computational Intelligence and Security, 2007;697-700.
36. Zetter K. Alleged 'SpyEye' Botmaster Ends Up in America, Handcuffs, Kim Zetter, Wired; 2013. Available:Wired.com. Retrieved 5 September, 2016
37. Saroiu S, Gribble S, Levy H. Measurement and analysis of spyware in a university environment. In Networked Systems

Design and Implementation (NSDI). 2004; 141-153.

38. Angiulli F, Fassetti F. Detecting Distance-Based Outliers in Streams of Data DEIS, Universit `a della Calabria Via P. Bucci, 41C 87036 Rende (CS), Italy; 2007. Available:f.angiulli@deis.unical.it. ffassetti@deis.unical.it